\documentclass[journal]{IEEEtran}
\usepackage{cite}
\usepackage{amsmath,amssymb,amsfonts}
\usepackage{graphicx}
\usepackage{textcomp}
\usepackage{xcolor}

\usepackage{hyperref}
\usepackage{amsmath}
\usepackage{autobreak}
\usepackage{amsfonts}
\usepackage{bm}
\usepackage{bbm}
\usepackage{algorithm}  
\usepackage{algorithmicx}  
\usepackage{algpseudocode}
\usepackage{stfloats}
\usepackage{cuted}
\usepackage{color}
\usepackage{subfigure}
\usepackage{cite}
\usepackage{booktabs} 

\newcommand{\st}{\text{s.t.}}
\newcommand{\diag}{\text{diag}}
\newcommand{\tr}{\text{Tr}}
\newcommand{\vect}{\text{vec}}

\makeatletter
\renewcommand{\maketag@@@}[1]{\hbox{\m@th\normalsize\normalfont#1}}%
\makeatother

\ifCLASSINFOpdf
\else
\fi
\begin{document}
	%
\title{Energy-Efficient Cell-Free Network \\ Assisted by Hybrid RISs}
	%
	%
	%
	
	\author{Wanting Lyu,~Yue Xiu,~Songjie Yang,~Chau Yuen,~\IEEEmembership{Fellow,~IEEE},~and~Zhongpei Zhang,~\IEEEmembership{Member,~IEEE} 
		\thanks{
			Wanting Lyu, Yue Xiu, Songjie Yang and Zhongpei Zhang are with National Key Laboratory of Science and Technology on Communications, University of Electronic Science and Technology of China, Chengdu61173, China (E-mail: lyuwanting@yeah.net; xiuyue12345678@163.com; yangsongjie@std.uestc.edu.cn; zhangzp@uestc.edu.cn).
			
			
			C. Yuen is with the Engineering Product Development (EPD) Pillar,
            Singapore University of Technology and Design, Singapore 487372 (e-mail:
            yuenchau@sutd.edu.sg).
    		}}

\maketitle

\begin{abstract}
In this letter, we investigate a cell-free network aided by hybrid reconfigurable intelligent surfaces (RISs), which consists of a mixture of passive and active elements that are capable of amplifying and reflecting the incident signal. To maximize the energy efficiency (EE) of the system, we formulate a joint transmit beamforming and RIS coefficients optimization problem. To deal with the fractional objective function, Dinkelbach transform, Lagrangian dual reformulation, and quadratic transform are utilized, with a block coordinate descent (BCD) based algorithm proposed to decouple the variables. In addition, successive convex approximation (SCA) method is applied to iteratively to tackle the non-convexity of the sub-problems. Simulation results illustrate the effectiveness and convergence of the proposed algorithm through analyzing the EE and sum rate performance with varying parameter settings. The proposed hybrid RISs schemes can achieve $92\%$ of the sum rate but $188\%$ of EE of active RISs schemes. As compared with passive RISs, $11\%$ gain in sum rate can be achieved with comparable EE.
\end{abstract}


\IEEEpeerreviewmaketitle

\section{Introduction}
\IEEEPARstart{C}{ell-free} network has gained great attractiveness in beyond fifth-generation mobile communications (B5G) without cell boundaries \cite{CFSurvey}. It has great potential in next generation indoor and hot-spot environment, such as shopping malls, train stations, hospitals and subways. In addition, cell-free network is particularly effective in high-mobility scenarios like vehicular networks without handover cost \cite{6Gmagzine}. Despite the above advantages, conventional cell-free network requires massive access points (APs), resulting in high power consumption both for AP transmit power and hardware power \cite{ZYT_CFris}. 

Recently, reconfigurable intelligent surface (RIS), as a promising technology to overcome obstacles, enhance channel capacity and improve energy efficiency \cite{Add2} in various scenarios \cite{YZH_EE,CY1-EEandSE,CY2-secure,JZJ_secret_key,CY4,Add1,Add3}, has been integrated into cell-free networks to replace some of the APs \cite{ZYT_CFris}\cite{CFperform_analysis2}\cite{DLL_CFris}. However, the ideal capacity gains provided by RIS is difficult to achieve practically owing to the "multiplicative fading" effect caused by RIS, causing extremely large path loss in the cascaded channels. Considering this, a new structure of RIS, namely active RIS has been proposed by authors of \cite{LYC_activeRIS}\cite{DLL_activeIRS}, where active reflecting elements are configured with radio frequency (RF) chains and power amplifiers to alleviate the severe fading effect. The active RIS, however, requires higher power consumption, and introduce non-negligible self-interference and thermal noise. Considering the trade-off between the signal amplifying effect and power consumption, a hybrid RIS architecture has been proposed \cite{HybridRIS_TVT}. In this architecture, only a few reflecting elements are activated, introducing a lower level of transmit power and effective noise than active RIS, while significantly enhancing the signal strength compared with full RIS. Hybrid RIS provides a reliable, sustainable solution to the wireless network design with an acceptable level of cost and power consumption. 

To the best of our knowledge, conventional passive RISs have been used in cell-free networks to reduce the power consumption, but the data rate is limited by the severe double-fading effect. To improve the energy efficiency (EE) of cell-free networks while alleviating the fading effect, we consider hybrid RISs with a few active elements capable of amplifying the incident signal to replace a part of APs. Thus, we consider a downlink cell-free network assisted by hybrid RISs to achieve a trade-off between EE and sum rate performance. A highly coupled non-convex EE maximization problem constraint on the minimum rate requirement is formulated by optimizing the digital beamforming and hybrid RIS coefficients design. Then, we propose a block coordinate descent (BCD) based algorithm to decouple the variables. To deal with the fractional objective function, Dinkelbach method, Lagrangian dual reformulation, and quadratic transform are applied. Particularly, successive convex approximation (SCA) method is used to tackle the non-convexity of the subproblems. Simulation results show the effectiveness of the proposed algorithm in terms of improving energy efficiency.
\begin{figure}[t]
	\centering
	\includegraphics[width=0.85\linewidth]{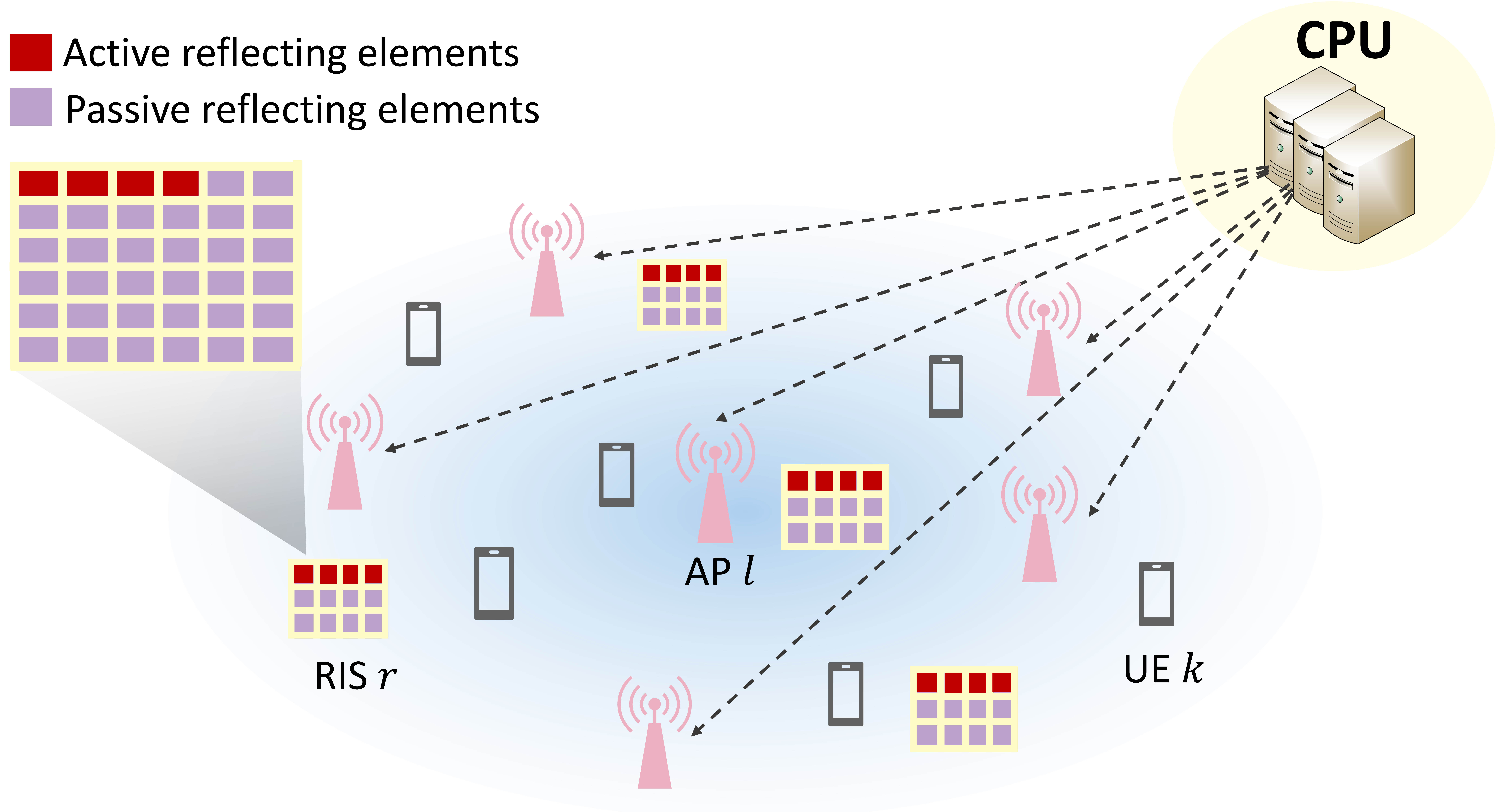}
	\caption{System model of the hybrid RIS-aided cell-free network.}
	\label{system model}
\end{figure}

\section{System Model and Problem Formulation}

\subsection{Transmission Model}
Consider a downlink cell-free network with $L$ APs each configured with $N_t$ antennas serving $K$ single-antenna users as shown in Fig. \ref{system model}. To enhance the communication quality and improve the energy efficiency, $R$ hybrid RISs are deployed in the network, each with $N_s$ reflecting elements. For each RIS, $N_a \ll N_s$ active reflecting elements are predetermined in the set $\mathcal A_r$, where $|\mathcal A_r| = N_a$. The RIS coefficient matrix of RIS $r$ is defined as $\mathbf \Theta_r = \text{diag}\{a_{r,1},\dots,a_{r,N_s}\}$, where $a_{r,n_s} = |a_{r,n_s}|e^{j\theta_{r,n_s}}$ denotes the coefficient of the $n_s^{th}$ element, with the amplitude $|a_{r,n_s}| \le a_{max}$ if the element $
(r,n_s)$ is in the set $\mathcal A_r$, and $|a_{r,n_s}| \le 1$ otherwise. To combine the elements in all $R$ RISs together and for ease of exposition, define $n = N_s(r-1) + n_s, 1\le n \le N$, where $N = RN_s$. The elements in set $\mathcal A = \mathop{\cup}\limits_r\mathcal A_r$ are the active reflecting elements in all $R$ RISs. Then, the overall RIS matrix can be written as $\mathbf \Theta = \text{blockdiag}\{\mathbf\Theta_1, \dots, \mathbf\Theta_R\}$. To distinguish the active and passive elements, rewrite $\mathbf\Theta$ as $\mathbf\Theta = \mathbf\Psi + \mathbf\Upsilon$, where $\mathbf\Psi = 
\mathbf A\mathbf\Theta$ and $\mathbf\Upsilon = (\mathbf I_N - \mathbf A)\mathbf\Theta$ denote the coefficient matrices of active and passive elements, respectively, where the selection matrix $\mathbf A$ is an $N\times N$ diagonal matrix with the elements corresponding to set $\mathcal A$ to be $1$, and others to be $0$.

Assume that all APs and all RISs serve all users over the same time-frequency resource, such that data symbol $\mathbf s \in \mathbb C^{K\times 1}$ satisfying $\mathbb E\{\mathbf{ss}^H\} = \mathbf I_K$ is transmitted from the APs simultaneously. A central processing unit (CPU) connected to all APs generates the digital beamformers based on perfectly known channel state information (CSI) of all the links. The transmitted signal at the AP $l$ can then be expressed as $\mathbf x_l = \mathbf W_l\mathbf s$, where $\mathbf W_l \in \mathbb C^{N_t\times K}$ denotes the transmit beamforming matrix at AP $l$. The links from AP $l$ to user $k$, from AP $l$ to RIS $r$, and from RIS $r$ to user $k$ are denoted as $\mathbf d_{l,k}^H \in \mathbb{C}^{1\times N_t}$, $\mathbf G_{l,r} \in \mathbb C^{N_s\times N_t} \text{ and } \mathbf f_{r,k}^H \in \mathbb C^{1\times N_s}$, respectively. Thus, the received signal of user $k$ can be expressed as 
\begin{equation}
	y_k = \sum_{l=1}^L(\mathbf d_{l,k}^H + \sum_{r=1}^R\mathbf f_{r,k}^H\mathbf\Theta_r\mathbf G_{l,r})\mathbf W_l\mathbf s + \sum_{r=1}^R \mathbf f_{r,k}^H\mathbf\Psi\mathbf z_r +  n_k,
	\label{Rx_sig}
\end{equation}
where $n_k \sim \mathcal{CN}(0,\sigma_0^2)$ is the additional white Gaussion noise (AWGN) at user $k$, and $\mathbf z_r \sim \mathcal{CN}(\mathbf 0, \sigma_r^2\mathbf A_r\mathbf I_{N_s})$ is the effective noise caused by the active elements of RIS $r$, including the AWGN and self-interference due to the full-duplex mode. Or we can write (\ref{Rx_sig}) in a more compact form as
\begin{equation}
	\begin{split}
		y_k & = (\mathbf d_k^H + \mathbf f_k^H\mathbf{\Theta G})\mathbf{Ws} + \mathbf f_k^H\mathbf{\Psi z} + n_k, \\
		& = \tilde{\mathbf h}_k^H\mathbf{Ws} + \mathbf f_k^H\mathbf{\Psi z} + n_k,
		\label{Rx_comb}
	\end{split}
\end{equation}
where $\mathbf d_k^H = (\mathbf d_{1,k}^H, \dots, \mathbf d_{L,k}^H) \in \mathbb C^{1\times LN_t}$, $\mathbf f_k^H = (\mathbf f_{1,k}^H, \dots, \mathbf f_{R,k}^H) \in \mathbb C^{1\times RN_s}$, $\mathbf W = (\mathbf W_1^T, \dots, \mathbf W_L^T)^T \in \mathbb C^{LN_t\times K}$, $\mathbf z = (\mathbf z_1,\dots,\mathbf z_R) \in \mathbb C^{RN_s\times 1}$, and 
\begin{equation}
	\mathbf G = \begin{pmatrix}
		\mathbf G_{1,1} & \cdots & \mathbf G_{L,1} \\
		\vdots         & \ddots & \vdots \\
		\mathbf G_{1,R} & \cdots & \mathbf G_{L,R}
	\end{pmatrix}
	\in \mathbb C^{RN_s\times LN_t}.
\end{equation}

Hence, the received signal-to-interference-plus-noise (SINR) at user $k$ can be expressed as 
\begin{equation}
	\gamma_k = \frac{|\tilde{\mathbf h}_k^H\mathbf w_k|^2}{\sum_{j\neq k}|\tilde{\mathbf h}_k^H\mathbf w_j|^2 + \sum_{r=1}^R\tilde{\sigma}_r^2 + \sigma_0^2},
	\label{Rx_SINR}
\end{equation}
where $\mathbf w_k$ is $k^{th}$ column of the digital beamforming matrix $\mathbf W$, and $\tilde{\sigma}_r^2 = \sigma_r^2||\mathbf f_{r,k}^H\mathbf\Psi_r||^2$ is the effective RIS noise power.

\subsection{Energy Efficiency}

The total power consumption includes transmit power at APs, transmit power at active elements of RISs and hardware circuit power. Therefore, the total power consumption of the cell-free system can be expressed as
\begin{align}
    P_{tot}(\mathbf W,\mathbf\Psi) &= \underbrace{\sum_{l=1}^LP_{l,tx}^A(\mathbf W_l)}_\text{transmit power of APs}  + \underbrace{\sum_{r=1}^RP_{r,a}^R(\mathbf\Psi_r,\mathbf W)}_\text{transmit power of active RIS elements} \notag\\ 
	& + \underbrace{ \sum_{l=1}^L P_{l,c}^A + \sum_{r=1}^RP_{r,c}^R + \sum_{k=1}^KP_k^U}_\text{circuit power of APs, RISs and UEs}.
		\label{P_total}
\end{align}
 Specifically, transmit power of AP $l$ is $P_{l,tx}^A(\mathbf W_l) = \frac{\text{Tr}(\mathbf W_l^H\mathbf W_l)}{\mu^A}$, and transmit power of the active elements of RIS $r$ is $P_{r,a}^R(\mathbf\Psi_r\mathbf W) = \frac{1}{\mu^R}\mathbb E\{||\mathbf\Psi_r(\mathbf G_r\mathbf W\mathbf s + \mathbf z_r)||^2\} =  \frac{1}{\mu^R}\left( \tr(\mathbf W^H\mathbf G_r^H\mathbf \Psi_r^H\mathbf\Psi_r\mathbf G_r\mathbf W) + \sum_{n_s\in\mathcal A_r}|a_{r,n_s}|^2\sigma_r^2\right)$, where $\mu^A, \mu^R \in (0,1]$ denote the amplifier efficiency factors of AP and RIS, respectively.

Thus, the energy efficiency of this system can be written as
\begin{equation}
	\eta = \frac{\sum_{k=1}^K R_k}{P_{tot}},
	\label{EE}
\end{equation}
where $R_k = \log_2(1+\gamma_k)$ is the data rate of user $k$.

\subsection{Problem Formulation}

To maximize the system energy efficiency, the optimization problem can be formulated as
\begin{subequations}
\begin{align}
	\text{(P1)}\; \max_{\mathbf W,\mathbf \Theta} \quad &\eta 
	\label{P1obj} \\
	\st\quad & P_{l,tx}^A(\mathbf W_l) \le P_{max}^A, \; \forall l\in\{1,\dots,L\},
	\label{Cons_TxPow} \\
	& 0 \le \theta_n \le 2\pi, \; \forall n,
	\label{Cons_angle} \\
	& |a_n| \le a_{max}, \; n\in\mathcal A
	\label{Cons_active} \\
	& |a_n| \le 1, \; n\notin \mathcal A
	\label{Cons_passive}  \\
	& R_k \ge R_\text{th}, \; \forall k,
	\label{Cons_rate}  \\
	& P_{r,a}^R(\mathbf\Psi_r,\mathbf W) \le P_{max}^R, \;\forall r,
	\label{Cons_RISpow} 
\end{align}
\end{subequations}
\noindent where (\ref{Cons_TxPow}), (\ref{Cons_RISpow}) are the transmit power constraints for AP $l$ and RIS $r$. (\ref{Cons_angle}), (\ref{Cons_active}) and (\ref{Cons_passive}) are the phase shift and amplitude constraints for all RIS elements, respectively.  $(\ref{Cons_rate})$ is to guarantee the data rate of user $k$.

However, problem (P1) is difficult to solve due to the strongly coupled variables, non-convex fractional objective function (\ref{P1obj}), and non-convex constraints (\ref{Cons_rate}). To solve this problem, we proposed an iterative BCD algorithm in section III.

\section{Joint Optimization Algorithm}

In this section, we first transform the complex fractional objective function into equivalent concave form, and propose a BCD based algorithm to jointly optimize beamforming design $\mathbf W$ at the APs and RIS coefficients $\mathbf \Theta$ at the RISs.

We first apply Dinkelbach's method in \cite{Dinkelbach}. By introducing an auxiliary variable $\hat y$, the objective function (\ref{P1obj}) can be transformed as 
\begin{equation}
	f_1(\mathbf W,\mathbf \Theta, \hat y) = \sum_{k=1}^K\log_2\left(1+ \gamma_k\right) - \hat y P_{tot}(\mathbf W,\mathbf \Psi), \label{Dink_objfunc}
\end{equation}
where $\hat y$ can be updated as 
\begin{equation}
	\hat{y}' = \frac{\sum_{k=1}^K\log_2\left(1+ \gamma_k\right)}{P_{tot}(\mathbf W,\mathbf \Psi)}
	\label{y_opt}
\end{equation}
in each iteration.

To further tackle the sum-logarithms in (\ref{Dink_objfunc}), Lagrangian dual reformulation in \cite{FP} is utilized, and thus $f_1(\mathbf W,\mathbf \Theta, \hat y)$ can be transformed by introducing a slack variable $\hat{\mathbf \epsilon} = [\hat\epsilon_1,...,\hat\epsilon_k]^T$:
\begin{equation}
	f_2(\mathbf W,\mathbf \Theta, \hat y,\hat{\mathbf\epsilon}) = \sum_{k=1}^K\log(1+\hat\epsilon_k) - \sum_{k=1}^K\hat\epsilon_k + \sum_{k=1}^K\frac{(1+\hat\epsilon_k)\gamma_k}{1+\gamma_k}.
	\label{f2}
\end{equation}
The optimal $\hat{\mathbf\epsilon}$ can be obtained by taking the first partial derivative of $f_2(\mathbf W,\mathbf \Theta, \hat y,\hat{\mathbf\epsilon})$ such that $\frac{\partial f_2(\mathbf W,\mathbf \Theta, \hat y,\hat{\mathbf\epsilon})}{\partial\hat\epsilon_k} = 0$, and then we can update $\hat{\mathbf\epsilon}$ as
\vspace*{-0.5\baselineskip}
\begin{equation}
	\hat\epsilon_k' = \gamma_k.
	\label{epsilon_opt}
\end{equation}

However, $\gamma_k$ is still a non-convex fraction, which can be decoupled by the quadratic transform \cite{FP}, and thus the objective function can be rewritten as (\ref{f3}) at the bottom of this page, where the optimal value of slack variable $\hat{\rho}_k$ can be determined by setting $\frac{\partial f_3(\mathbf W,\mathbf \Theta, \hat y,\hat{\mathbf\epsilon},\hat{\mathbf\rho})}{\partial \rho_k} = 0$ such that \begin{figure*}[b]
	\hrule
	\begin{equation}
			f_3(\mathbf W,\mathbf \Theta, \hat y,\hat{\mathbf\epsilon},\hat{\mathbf\rho})  = \sum_{k=1}^K 2\sqrt{1+\hat\epsilon_k}\mathcal Re\left\{\left(\hat\rho_k^*\tilde{\mathbf h}_k^H\mathbf w_k\right)\right\}  -|\hat\rho_k|^2\left( \sum_{j=1}^K\left|\tilde{\mathbf h}_k^H\mathbf w_j\right|^2 + \sum_{r=1}^R\tilde{\sigma}_r^2 + \sigma_0^2\right) -\hat yP_{tot}(\mathbf{W,\Psi})
			\label{f3}
	\end{equation}
	\begin{equation}
		\begin{split}
			&f_4(\mathbf\theta | \mathbf W',\hat y', \hat\epsilon',\hat\rho' ) = \sum_{k=1}^K2\sqrt{1+\hat\epsilon_k'}\,\mathcal Re\Big\{ \hat\rho_k'^{*}(\mathbf d_k^H + 
			\mathbf\theta^H\diag\{\mathbf f_k^H\mathbf G\})\mathbf w_k' \Big\}  \\
			- &|\hat\rho_k|^2\Big(\sum_{k=1}^K\mathbf\theta^H Q_{k,j}^1\mathbf\theta + 2\mathcal Re\{\mathbf\theta^H Q_{k,j}^2\} 
			+ Q_{k,j}^3 + \sum_{r=1}^R\tilde{\sigma}_r'^{2} + \sigma_0^2 \Big) - \hat y'\Big( \sum_{r=1}^R P_{r,a}(\mathbf\theta_r|\mathbf W') 
			+ \sum_{l=1}^LP_{l,tx}^A(\mathbf W_l') + P_{hw} \Big)
		\end{split}
		\label{f4}\tag{25}
	\end{equation}
\end{figure*}
\begin{equation}
	\hat\rho_k' = \frac{\sqrt{1+\hat\epsilon_k}\tilde{\mathbf h}_k^H\mathbf w_k }{\sum_{j=1}^K|\tilde{\mathbf h}_k^H\mathbf w_j|^2 + \sum_{r=1}^R\tilde{\sigma}_r^2 + \sigma_0^2}.
	\label{rho_opt}
\end{equation}

To this end, problem (P1) is equivalent to
\begin{subequations}
\begin{align}
	\text{(P2)}\quad \max_{\mathbf W,\mathbf \Theta} \quad & f_3(\mathbf W,\mathbf \Theta, \hat y,\hat{\mathbf\epsilon},\hat{\mathbf\rho}) 
	\label{P2obj} \\
	\st\quad &  (\ref{Cons_TxPow}),(\ref{Cons_angle}),(\ref{Cons_active}),(\ref{Cons_passive}),(\ref{Cons_rate}),(\ref{Cons_RISpow}), \notag
\end{align}
\end{subequations}
which is still non-convex due to the coupling between variables $\mathbf W$ and $\mathbf\Theta$, and the non-convex fractional constraint (\ref{Cons_rate}). To decouple the variables, BCD method is applied to update the variables $\hat y,\ \hat{\mathbf\epsilon},\ \hat{\mathbf\rho}$, $\mathbf W$ and $\mathbf\Theta$ alternatively.

\subsection{Transmit Beamforming Optimization}

Given fixed $\mathbf\Theta',\hat y',\hat{\mathbf\epsilon}'$ and $\hat{\mathbf\rho}'$ obtained in the last iteration, (P2) can be updated as
\begin{subequations}
\begin{align}
	\text{(P3)}\quad \max_{\mathbf W} \quad & f_3(\mathbf W|\mathbf \Theta', \hat y',\hat{\mathbf\epsilon}',\hat{\mathbf\rho}') 
	\label{P3obj} \\
	\st\quad &  (\ref{Cons_TxPow}),(\ref{Cons_rate}),(\ref{Cons_RISpow}). \notag
\end{align}
\end{subequations}

Note that the objective function is concave with respect to $\mathbf W$. Constraints (\ref{Cons_TxPow}) and (\ref{Cons_RISpow}) can be transformed as
\begin{equation}
		P_{l,tx}^A(\mathbf W_l) \overset{\text{(a)}}{=} \frac{1}{\mu^A}\vect\left((\mathbf W_l)\right)^H\vect\left(\mathbf W_l\right) \text{ and } \label{AP_Tx_power_kron}
\end{equation}
\begin{align}
	&P_{r,a}^R(\mathbf W|\mathbf \Psi_r^{(t)}) \overset{\text{(b)}}{=}  \frac{1}{\mu^R}\bigg(\sum_{[r,n_s]\in\mathcal A_r}|a_{r,n_s}^{(t)}|^2\sigma_r^2 \notag\\
	&+\big(\vect\left(\mathbf W\right)\big)^H \left(\mathbf I\otimes \left(\mathbf G_r^H\mathbf\Psi_r^{(t)H}\mathbf\Psi_r^{(t)}\mathbf G_r\right)\vect(\mathbf W)\right) \bigg).
    \label{RIS_Tx_power_kron}
\end{align}
\noindent Note that (a), (b) are because of $\tr(\mathbf A^H\mathbf B) = (\vect(\mathbf A))^H\vect(\mathbf B)$ and $\tr(\mathbf{ABC}) = (\vect(\mathbf A^H))^H(\mathbf I\otimes \mathbf B)\vect(\mathbf C)$, respectively. 

Moreover, the non-convex constraint (\ref{Cons_rate}) can be rewritten as
\begin{equation}
	|\tilde{\mathbf h}_k^H\mathbf w_k|^2 \ge \gamma_\text{th}\left({\sum_{j\neq k}|\tilde{\mathbf h}_k^H\mathbf w_j|^2 + \sum_{r=1}^R\tilde{\sigma}_r^2 + \sigma_0^2}\right) 
	\label{Cons_SINR}
\end{equation}
where $\gamma_\text{th} = 2^{R_\text{th}}-1$. Applying SCA method, the left hand side of the inequality can be lower bounded by its first Taylor expansion, and thus (\ref{Cons_SINR}) can be approximated as
\begin{equation}
	\begin{split}
		& 2\mathcal Re\left\{ \left(\tilde{\mathbf h}_k^H\mathbf w_k^{(i)}\right)^H\tilde{\mathbf h}_k^H\mathbf w_k \right\} - |\tilde{\mathbf h}_k^H\mathbf w_k^{(i)}|^2 \ge \\
		&\gamma_\text{th}\left({\sum_{j\neq k}|\tilde{\mathbf h}_k^H\mathbf w_j|^2 + \sum_{r=1}^R\tilde{\sigma}_r^2 + \sigma_0^2}\right),
		\label{Taylor_W}
	\end{split}
\end{equation}
where $(i)$ denotes the value in the last iteration. Thus, (P3) can be reformulated as
\begin{subequations}
\begin{align}
	\text{(P3-1)}\quad \max_{\mathbf W} \quad & f_3(\mathbf W|\mathbf \Theta', \hat y',\hat{\mathbf\epsilon}',\hat{\mathbf\rho}') 
	\label{P3obj} \\
	\st\quad & (\ref{AP_Tx_power_kron}), (\ref{RIS_Tx_power_kron}), (\ref{Taylor_W}), \notag 
\end{align}
\end{subequations}
which is a convex problem that can be efficiently solved by standard convex solvers such as CVX \cite{boyd2004convex}.

\subsection{Hybrid RIS Coefficients Optimization}

Given fixed $\mathbf W',\hat y',\hat{\mathbf\epsilon}'$ and $\hat{\mathbf\rho}'$, (P2) becomes
\begin{subequations}
\begin{align}
	\text{(P4)}\quad \max_{\mathbf \Theta} \quad & f_3(\mathbf \Theta, |\mathbf W',\hat y',\hat{\mathbf\epsilon}',\hat{\mathbf\rho}') 
	\label{P4obj} \\
	\st\quad &  (\ref{Cons_active}),(\ref{Cons_passive}),(\ref{Cons_rate}),(\ref{Cons_RISpow}). \notag
\end{align}
\end{subequations}

To make it more tractable, we first define a coefficient vector of RIS $r$ as $\mathbf\theta_r = (a_{r,1}, ..., a_{r,N_s})^H$. Accordingly, the active coefficient vector of RIS $r$ can be represented as $\mathbf\psi_r = \mathbf A_r\mathbf\theta_r$, where $\mathbf A_r$ denotes the active element selection matrix for RIS $r$. Thus, the equivalent channel can be rewritten as $\tilde{\mathbf h}_k^H = \mathbf d_k^H + \sum_{r=1}^R\mathbf\theta_r^H\diag\{\mathbf f_{r,k}^H\}\mathbf G_r = \mathbf d_k^H + \mathbf\theta^H\diag\{\mathbf f_k^H\}\mathbf G$, and $\mathbf\theta = (\mathbf\theta_1^T, ... , \mathbf\theta_R^T)^T$ denotes the overall RIS coefficient vector. Then, the transmit power of RIS $r$ can be rewritten as 
\begin{align}
		& P_{r,a}(\mathbf\theta_r|\mathbf W') \notag\\
		=\;& \frac{1}{\mu^R}\left(\tr\left( \mathbf\Psi_r\mathbf G_r\mathbf W'\mathbf W'^{H}\mathbf G_r^H\mathbf\Psi_r^H \right) + \sum_{n_s\in\mathcal{A}_r}\sigma_r^2\left|a_{r,n_s}\right|^2 \right) \notag\\
		=\;& \frac{1}{\mu^R}\left( \mathbf\psi_r^H\mathbf H_r\mathbf\psi_r + \sigma_r^2\mathbf\psi_r^H\mathbf\psi_r \right) \notag\\
		=\; & \frac{1}{\mu^R}(\mathbf\theta_r^H\mathbf A_r\mathbf H_r\mathbf A_r\mathbf\theta_r + \sigma_r^2\mathbf\theta_r^H\mathbf A_r\mathbf\theta_r),
\end{align}
where $\mathbf H_r$ is a diagonal matrix with its diagonal elements equal to those of the Hermitian matrix $\mathbf G_r\mathbf W'\mathbf W'^{H}\mathbf G_r^H$. Besides, the effective RIS noise power can be rewritten as $\tilde{\sigma}_{r,k}'^{2} = \sigma_r^2||\diag\{\mathbf f_{r,k}^H\}\mathbf{A_r\theta_r}||^2$. Similar as the last subsection, the non-convex constraint (\ref{Cons_rate}) is first rewritten as (\ref{Cons_SINR}), and then further transformed as
\begin{small}
	\begin{equation}
		\begin{split}
			&\mathbf\theta^HQ_{k,k}^1\mathbf\theta + 2\mathcal Re\{\mathbf\theta^HQ_{k,k}^2\} + Q_{k,k}^3 \\
			\ge&\  \gamma_\text{th}\Bigg( \sum_{j\neq  k}\mathbf\theta^HQ_{k,j}^1\mathbf\theta   
			+ 2\mathcal Re\{\mathbf\theta^HQ_{k,j}^2\} + Q_{k,j}^3 + \sum_{r=1}^R\tilde{\sigma}_{r,k}'^{2} + \sigma_0^2 \Bigg),
			\label{SINR_Q}
		\end{split}
	\end{equation}
\end{small}%
where $Q_{k,j}^1 = \diag\{\mathbf f_k^H\}\mathbf G\mathbf w_j'\mathbf w_j'^{H}\mathbf G^H\diag\{\mathbf f_k\}$, $Q_{k,j}^2 = \diag\{\mathbf f_k^H\}\mathbf G\mathbf w_j'\mathbf w_j'^{H}\mathbf d_k$ and $Q_{k,j}^3 = |\mathbf d_k^H\mathbf w_j'|^2$, respectively. SCA method can be applied again to approximate (\ref{SINR_Q}) as
\begin{small}
	\begin{equation}
		\begin{split}
			&2\mathcal Re\{\mathbf\theta^{(i)H}Q_{k,k}^1\mathbf\theta\} - \mathbf\theta^{(i)H}Q_{k,k}^1\mathbf\theta^{(i)} + 2\mathcal Re\{\mathbf\theta^HQ_{k,k}^2\} + Q_{k,k}^3  \\
			\ge&\  \gamma_\text{th}\Bigg( \sum_{j\neq  k}\mathbf\theta^HQ_{k,j}^1\mathbf\theta  
			+ 2\mathcal Re\{\mathbf\theta^HQ_{k,j}^2\} + Q_{k,j}^3 + \sum_{r=1}^R\tilde{\sigma}_{r,k}'^{2} + \sigma_0^2 \Bigg),
			\label{Cons_SINR_theta}
		\end{split}
	\end{equation}
\end{small}%
where $\mathbf\theta^{(i)}$ denotes the optimal value of $\mathbf\theta$ in the last iteration.

To explicitly represent the variable $\mathbf\theta$, the objective function becomes (\ref{f4}) at the bottom of the last page ($P_{hw}$ denotes the overall circuit power of the system), which is concave with respect to $\mathbf\theta$.

As a result, (P4) is reformulated as
\begin{subequations}
\begin{align}
	\text{(P4-1)}\quad \max_{\mathbf \Theta} \quad & f_4(\mathbf \Theta, |\mathbf W',\hat y',\hat{\mathbf\epsilon}',\hat{\mathbf\rho}') 
	\label{P4-1obj} \\
	\st\quad &  (\ref{Cons_active}),(\ref{Cons_passive}),(\ref{Cons_SINR_theta}), \notag \\
	& P_{r,a}(\mathbf\theta_r|\mathbf W') \le P_{max}^R. \label{Cons_RISpow_theta} \tag{\ref{P4-1obj}{a}}
\end{align}
\end{subequations}
Note that in each iteration, (P4-1) is a convex problem that can be easily solved by existing methods. 

\subsection{Overall Algorithm and Complexity Analysis}

The overall BCD based algorithm is summarized in \textbf{Algorithm \ref{Algorithm1}}, where in each iteration, variables $\hat y,\hat{\mathbf\epsilon},\hat{\mathbf\rho}, \mathbf W$ and $\mathbf\theta$ are updated step by step.
\begin{algorithm}[t]  
	\caption{Proposed Algorithm for the EE Maximization Problem}
	\begin{algorithmic}[1]  
		\State \textbf{Initialize} variables $ \hat y^{(0)},\hat{\mathbf\epsilon}^{(0)},\hat{\mathbf\rho}^{(0)}, \mathbf W^{(0)}$ and $\mathbf\theta^{(0)}$. Set iteration index $i=1$.
		\Repeat 
			\State Update Dinkelbach slack variable $\hat y'^{(i)}$ by (\ref{y_opt}).
			\State Update Lagrangian dual reformulation slack variable $\hat{\mathbf\epsilon}'^{(i)}$ by (\ref{epsilon_opt}).
			\State Update quadratic transform slack variable $\hat{\mathbf\rho}'^{(i)}$ by (\ref{rho_opt}).
			\State Update transmit beamforming $\mathbf W'^{(i)}$ by solving (P3-1).
			\State Update RIS coefficients $\mathbf\theta'^{(i)}$ by solving (P4-1).
			\State Update EE $\eta$.
		\Until{convergence of $\eta$.}
	\end{algorithmic} 
	\label{Algorithm1}
\end{algorithm}

The computational complexity of the overall algorithm consists of the complexity for updating the variables $\hat y,\hat{\mathbf\epsilon},\hat{\mathbf\rho}, \mathbf W$ and $\mathbf\theta$, which mainly depends on the iterative algorithm for solving $\mathbf W$ and $\mathbf\theta$. Thus, the overall computation complexity of the proposed algorithm is $\mathcal O(I_0(K^{3.5}\log_2(1+\epsilon_1) + N^{3.5}\log_2(1+\epsilon_2) ))$, where $I_0$ is the number of iterations. Besides, $\epsilon_1$ and $\epsilon_2$ denote the accuracy of SCA algorithms.


\section{Numerical Results}

In this section, we provide the numerical simulation results to show the effectiveness of our proposed algorithm for the hybrid-RIS aided cell-free network and compare the performance with other benchmarks.

Without loss of generality, we consider a  $\rm{200 m \times 200 m}$ square area. Specifically, $L = 4$ APs (each with $N_t = 6$) transmit antennas and $R = 2$ RISs are distributedly deployed in the area serving $K = 4$ single-antenna users simultaneously. Rician fading model is used in each link, with the Rician factor to be $\beta = 3\text{ dB}$. The path loss (PL) of each link is $L = C_0(d/d_0)^{-\alpha}$, where $C_0 = -30 \text{ dB}$ is the reference PL at $d_0 = 1\text{ m}$, $d$ is the length of the corresponding links, and $\alpha = \{2.8,2.2,2.2\}$ are the path loss exponents of AP-UE, AP-RIS and RIS-UE links, respectively. Assume that the first $N_a$ elements of each RIS are active, while others are passive. The equivalent noise power of active RIS elements are set as $\sigma_r^2 = -76\text{ dBm}$, and the AWGN power at the users are $\sigma_0^2 = -80 \text{ dBm}$. The amplifier efficiencies are $\mu^A = \mu^R = 0.8$, and the minimum rate limit is $R_\text{th} = 0\text{ dB}$.

\begin{figure*}[t]
    \centering
	\subfigure[Energy efficiency versus transmit power at APs ]{
		\begin{minipage}[t]{0.31\linewidth}
			\centering	\includegraphics[width=\linewidth]{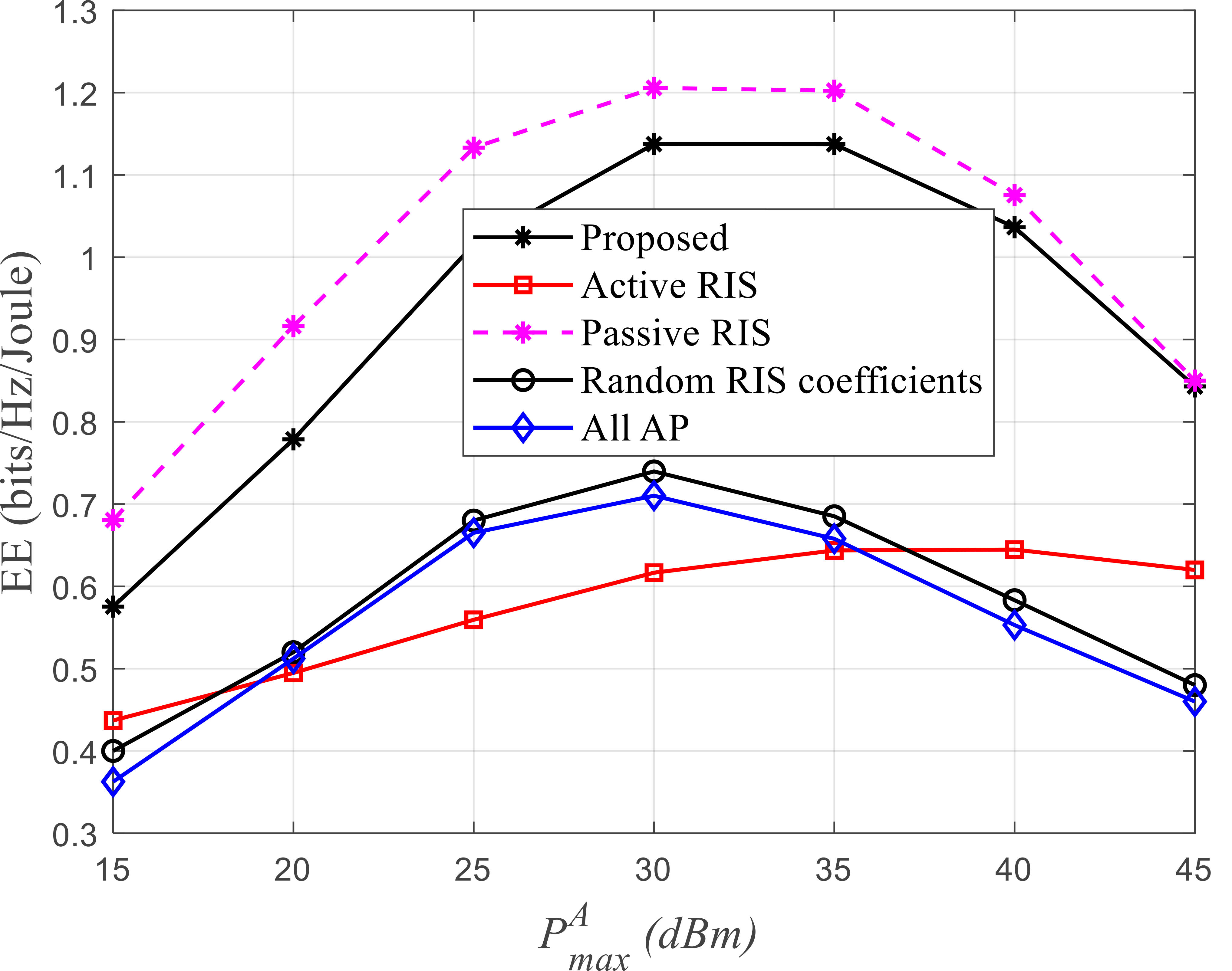}
        	\label{EEvsPa}
		\end{minipage}
	}
	\subfigure[Sum rate versus transmit power at APs]{
		\begin{minipage}[t]{0.31\linewidth}
			\centering
    	\includegraphics[width=\linewidth]{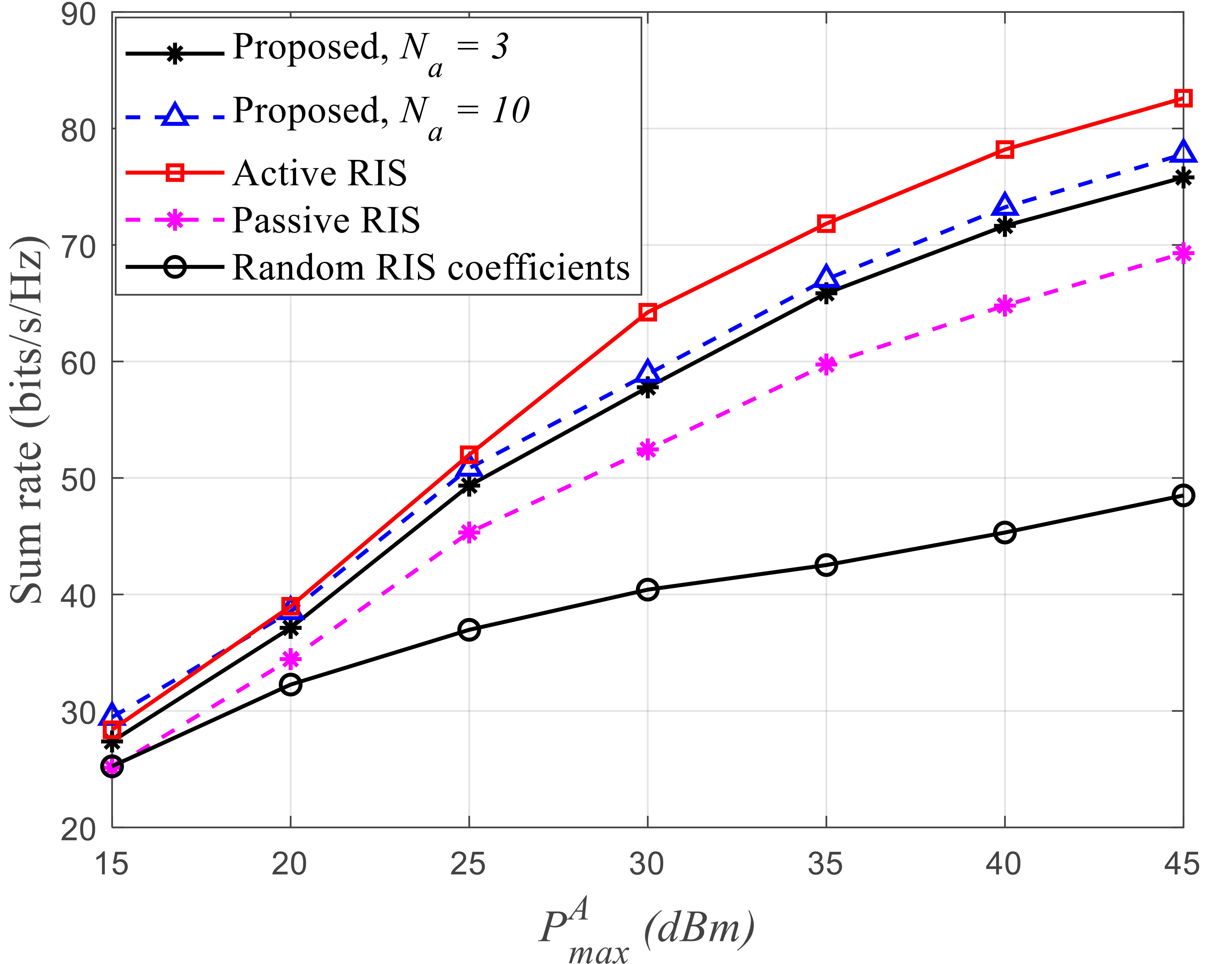}
    	\label{SumRatevsPa}
		\end{minipage}
	}
	\subfigure[Energy efficiency versus the number of iterations]{
		\begin{minipage}[t]{0.305\linewidth}
			\centering
        	\includegraphics[width=\linewidth]{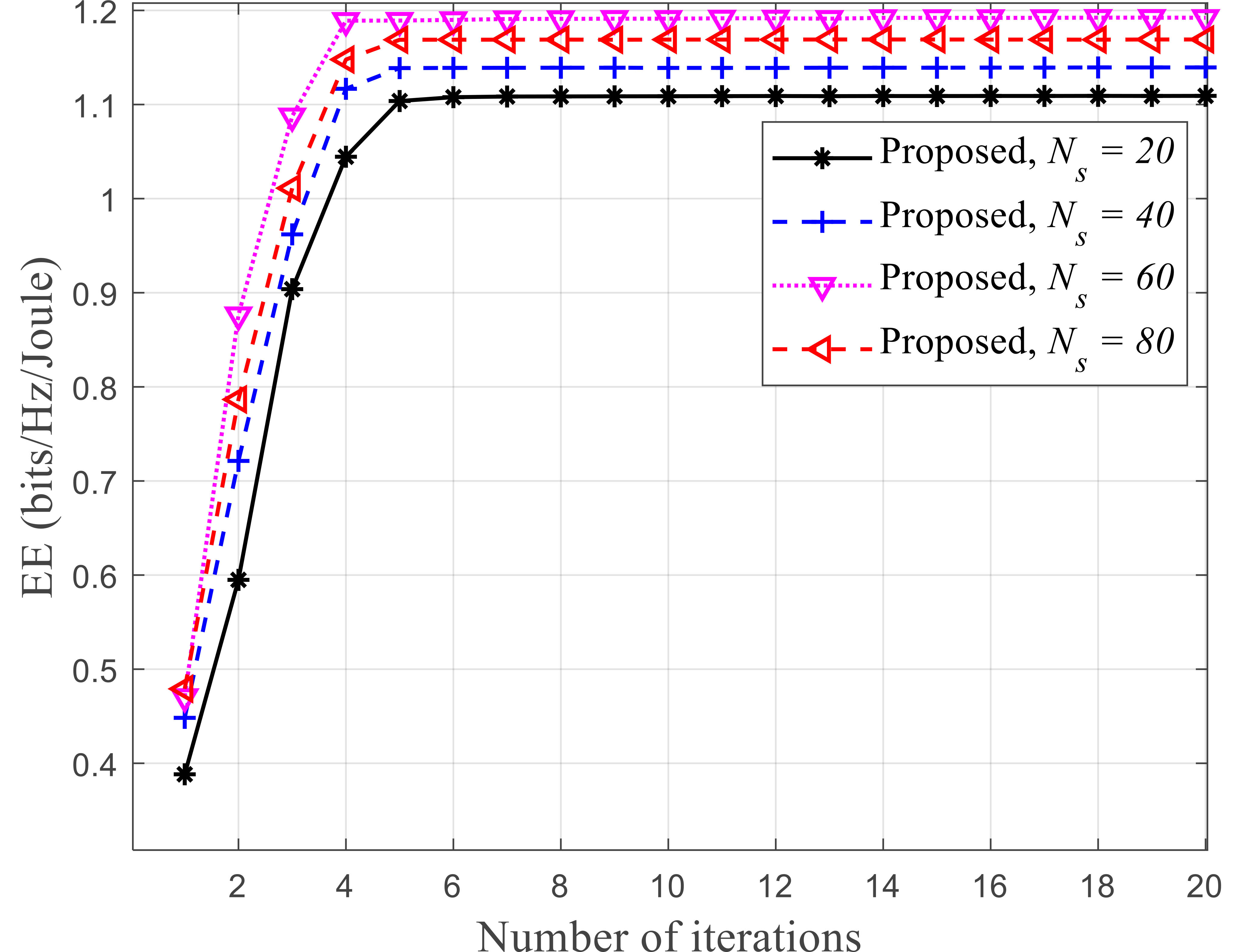}
        	\label{EEvsIte}
		\end{minipage}
	}
	\caption{Simulation results}
\end{figure*}

To begin with, the EE performance is analyzed with increasing transmit power at the APs illustrated in Fig. \ref{EEvsPa}. We consider each RIS are configured with $N_s = 80$ elements including $N_a = 3$ active ones. The maximum RIS transmit power is assumed as $P_{max}^R = 10 \text{ dBm}$. We can see from the figure that the EE of the system first grows with transmit power AP increases, and then drops for each configuration. The proposed algorithm for the hybrid RIS significantly outperforms the active RIS ($N_a = N_s$), random RIS coefficients and all AP cases, where all AP means to replace the RISs with APs. For fairness, the overall transmit power keeps unchanged. In addition, the EE performance of the proposed algorithm with hybrid RISs approaches about $96\%$ of that with conventional fully passive RISs with the AP transmit power increasing.

We further investigate the sum rate performance of the proposed system versus transmit power at the APs. From Fig. \ref{SumRatevsPa}, the system achievable sum rate is keep increasing with higher AP transmit power budget. Specifically, active RISs aided cell-free networks achieves better sum rate than other cases. The hybrid RISs with $N_a = 10$ active elements slightly performs better than those with only $N_a = 3$ active elements. The effectiveness of the proposed algorithm can be seen compared with passive RIS and random RIS coefficients cases. Combined with results in Fig. \ref{EEvsPa}, it can be inferred that the hybrid RIS takes into account a trade-off between EE and sum rate performance, compared with active and passive RISs.

The convergence of the proposed algorithm is verified in Fig. \ref{EEvsIte}, from which we can see the results converges fast within around $5$ iterations. Influence of different numbers of RIS elements $N_s$ is analyzed as well. By increasing $N_s$ from $20$ to $60$, the EE performance is improved due to the diversity gain of more RIS elements. Then the system EE drops slightly with $N_s = 80$, since a larger scale of RIS has higher circuit power consumption.

\section{Conclusion}

In this letter, we studied a hybrid RISs assisted cell-free network. The energy efficiency maximization problem was formulated by jointly optimizing the transmit beamforming and hybrid RIS coefficients design. We first transform the coupled fractional objective function into a concave form by introducing slack variables. Then, a BCD based iterative algorithm was proposed to solve the non-convex problem with SCA method. The simulation results showed that the proposed hybrid RISs algorithm can achieve comparable sum rate as active RISs but as energy-efficient as passive RISs. Also, the fast convergence was verified by the numerical results as well.

\ifCLASSOPTIONcaptionsoff
\newpage
\fi

\bibliographystyle{IEEEtran}
\bibliography{Ref}

\end{document}